\documentclass[sigconf]{acmart}
\AtBeginDocument{%
  }

\setcopyright{acmlicensed}
\copyrightyear{2025}
\acmYear{2025}
\acmDOI{XXXXXXX.XXXXXXX}
\acmConference[UIST '25]{The 37th Annual ACM Symposium on User Interface Software and Technology}{September 28-–October 1, 2025}{Busan, Republic of Korea}
\acmISBN{978-1-4503-XXXX-X/2018/06}

\usepackage{graphicx}
\begin{document}

%%
%% The "title" command has an optional parameter,
%% allowing the author to define a "short title" to be used in page headers.
\title{DESAMO: A Device for Elder-Friendly Smart Homes Powered by Embedded LLM with Audio Modality}

%%
%% The "author" command and its associated commands are used to define
%% the authors and their affiliations.
%% Of note is the shared affiliation of the first two authors, and the
%% "authornote" and "authornotemark" commands
%% used to denote shared contribution to the research.
\author{Youngwon Choi}
\authornote{Both authors contributed equally to this research. Youngwon Choi led the on-device execution of Audio LLM architecture and designed the prompting strategies. Donghyuk Jung developed the interface components and contributed to overall system integration.}
\affiliation{%
  \institution{MAUM AI Inc.}
  \city{Seongnam}
  \country{Republic of Korea}
}
\email{youngwonchoi@maum.ai}

\author{Donghyuk Jung}
\authornotemark[1]
\affiliation{%
  \institution{Korea Culture Technology Institute}
  \city{Gwangju}
  \country{Republic of Korea}
}
\email{dhjung081121@gm.gist.ac.kr}

\author{Hwayeon Kim}
\affiliation{%
  \institution{MAUM AI Inc.}
  \city{Seongnam}
  \country{Republic of Korea}
}
\email{khy0908@maum.ai}

%%
%% By default, the full list of authors will be used in the page
%% headers. Often, this list is too long, and will overlap
%% other information printed in the page headers. This command allows
%% the author to define a more concise list
%% of authors' names for this purpose.
\renewcommand{\shortauthors}{Trovato et al.}

%%
%% The abstract is a short summary of the work to be presented in the
%% article.
\begin{abstract}
    We present DESAMO, an on-device smart home system for elder-friendly use powered by Audio LLM, that supports natural and private interactions. While conventional voice assistants rely on ASR-based pipelines or ASR–LLM cascades, often struggling with the unclear speech common among elderly users and unable to handle non-speech audio, DESAMO leverages an Audio LLM to process raw audio input directly, enabling a robust understanding of user intent and critical events, such as falls or calls for help. 
    % To our knowledge, DESAMO is the first academic system to utilize an Audio LLM operating entirely on an edge device, offering a unified and privacy-preserving interface for aging-in-place environments.

\end{abstract}

%%
%% The code below is generated by the tool at http://dl.acm.org/ccs.cfm.
%% Please copy and paste the code instead of the example below.
%%


%%
%% Keywords. The author(s) should pick words that accurately describe
%% the work being presented. Separate the keywords with commas.
%% A "teaser" image appears between the author and affiliation
%% information and the body of the document, and typically spans the
%% page.
\begin{teaserfigure}
  \includegraphics[width=1\textwidth]{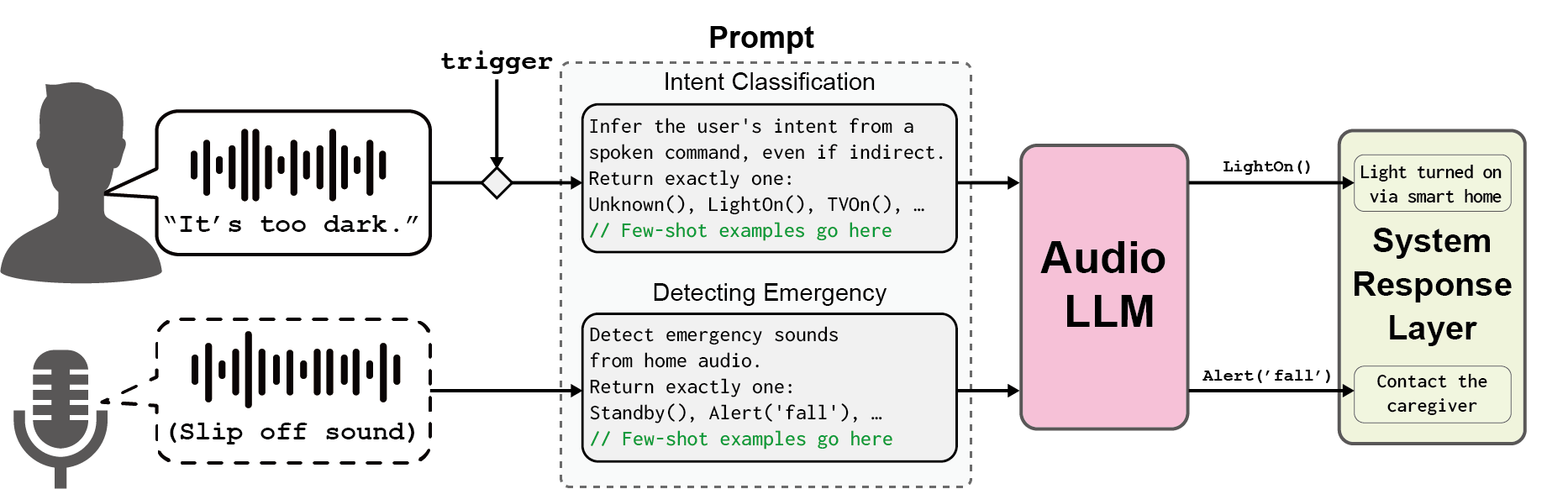}
  \caption{Pipeline Overview of DESAMO}
  \Description{Pipeline Overview of DESAMO}
  \label{fig:teaser}
\end{teaserfigure}

% \received{20 February 2007}
% \received[revised]{12 March 2009}
% \received[accepted]{5 June 2009}

%%
%% This command processes the author, affiliation, and title
 %% information and builds the first part of the formatted document..
\maketitle

\section{Introduction}
With the rise of aging populations and single-person households, there has been a steady increase in demand for smart home technologies that support personalized and contactless interaction~\cite{dawood2024management}. Voice assistant interfaces~\cite{guaman2018device, yue2017voice}, in particular, have become an intuitive means of controlling devices without manual input. However, conventional voice assistant systems typically rely on automatic speech recognition (ASR) based transcription followed by shallow intent parsers, making them poorly suited to interpret indirect expressions and ambiguous language \cite{mitra2019leveraging}. To address these limitations, recent systems have introduced ASR–LLM pipelines \cite{chen2024voicebench}, in which transcribed speech is passed to large language models. Nevertheless, these systems remain vulnerable to transcription errors~\cite{everson2024towards}, particularly when processing the unclear articulation of elderly users~\cite{vipperla2008longitudinal}, and such systems are not capable of processing non-speech audio inputs. More recently, large language models have been extended to the audio modality (Audio LLM), such as AudioChatLLaMA \cite{fathullah2024audiochatllama} and Qwen-Audio \cite{chu2023qwen}, enabling end-to-end inference directly from both speech and non-speech audio inputs. By jointly reasoning over acoustic and semantic information, these models allow a more robust and flexible audio understanding, eliminating the need for intermediate text representations. 

In this work, we propose DESAMO, an audio-interactive smart home system for older adults based on the Audio LLM that runs entirely on-device. DESAMO is capable of understanding not only direct and indirect spoken instructions, but also environmental sounds such as falls and screams, allowing for both voice intent classification and emergency event detection within a single unified model. Figure~\ref{fig:teaser} presents the overall architecture that supports both tasks using a shared Audio LLM. By executing all computations locally on edge hardware, this design inherently enhances user privacy, which is widely recognized as a critical factor in the adoption of smart homes~\cite{schomakers2020privacy}. To the best of our knowledge, this represents one of the first academic demonstrations of a fully on-device Audio LLM implemented as a functional prototype. 

\begin{table}[t]
\centering
\captionsetup{skip=4pt}
\caption{Performance of Cascaded and Proposed Approaches}
\label{tab:performance}
\begin{tabular}{lccc}
\toprule
\textbf{Model} & \textbf{WER} & \textbf{Accuracy} & \textbf{Model Size} \\
\midrule
Whisper medium + LLM     & 4.55\% & 96.33\% & 3.64GB \\
Whisper large-v3 + LLM   & 3.89\% & 97.33\% & 5.20GB \\
\textbf{Proposed}        & N/A    & \textbf{98\%} & \textbf{3.45GB} \\
\bottomrule
\end{tabular}
\end{table}

\section{DESAMO Prototyping}

\subsection{On-device DESAMO Execution}

DESAMO is built upon the Qwen2.5-Omni 3B \cite{xu2025qwen2}, a recent multimodal language model capable of processing both speech and ambient audio. For audio understanding, this model uses a Whisper large-v3 based audio encoder~\cite{radford2023robust} to transform the raw wave file into semantic embeddings, which are then fed into the language model. The entire system runs locally on NVIDIA Jetson Orin Nano, enabling full inference on-device without any cloud dependency. We use a quantized model with 16-bit audio encoders and a 4-bit language model, packaged in the compact GGUF format.

\subsection{Voice Intent Classification}

Recent advancements in function calling have enabled language models to generate structured function representations from natural language queries. Prior works \cite{schick2023toolformer,chen2024octopus, patil2024gorilla} demonstrate that even 2B to 7B scale models can effectively retrieve and generate functions based on user queries. In our use case, intent classification can be interpreted as a form of function calling, where natural language inputs like “call my daughter” are mapped to structured commands such as \textit{Call(`daughter')}. Inspired by this paradigm, DESAMO extends function calling to the audio domain, allowing users to issue commands using natural speech rather than text-based input.

Upon detecting a trigger phrase, DESAMO records a short audio segment that may contains either a direct command like "turn on the air conditioner" or an indirect expression such as "It's getting hot." The speech is processed into a semantic embedding and passed into the intent classification pipeline, along with a prompt that guides the model to generate structured control outputs such as \textit{ACOn()}. The system response layer interprets this output to activate the corresponding device, accompanied by a brief voice confirmation.

\subsection{Detecting Emergencies}

To ensure safety in environments where aging is an issue, DESAMO uses passive audio monitoring to detect emergencies, such as falls or distress, drawing inspiration from recent advances in prompt-based sound understanding \cite{ghosh2024gama, tang2024extending} with Audio LLM.

\begin{figure}[!htbp]
    \captionsetup{skip=10pt}
    \centering    \includegraphics[width=0.8\linewidth]{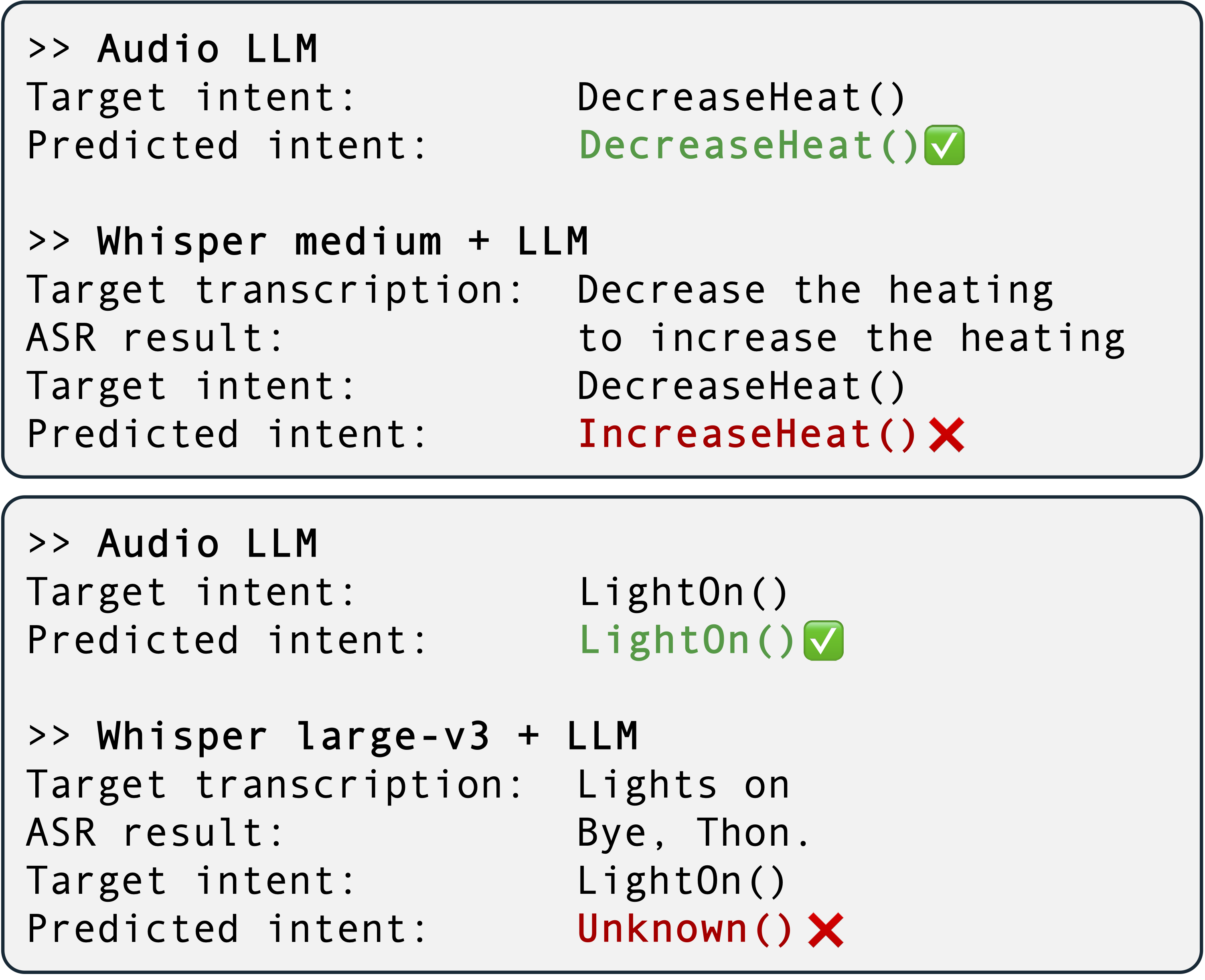} % change file name as needed
    \caption{Audio LLM directly predicts intent without suffering from ASR error propagation.}
    \Description{}
    \label{fig:result}
\end{figure}

The system continuously monitors ambient audio by capturing short sound segments at regular intervals. At each interval, the system feeds an audio segment and a detection prompt into the Audio LLM to identify potential emergencies, such as falls or verbal cries for help, from elderly users. The model outputs structured event labels, such as \textit{Alert(`fall')} or \textit{Alert(`help')}, which are interpreted by the system response layer to trigger responses, including sending an alert or sounding an alarm, and notifying caregivers.

\section{Pilot Evaluation}

To evaluate DESAMO in a realistic use case, we built a pilot benchmark using 300 samples curated from the Fluent Speech Commands dataset \cite{lugosch2019speech}, filtering for speakers aged 65 years or older, and selecting to match the voice intent classification scenario of DESAMO. We compared DESAMO with two cascaded baselines: one combining Whisper large-v3 with Qwen2.5-Omni in text-only mode, and another using Whisper medium with the same LLM. Whisper medium was selected because its parameter size is comparable to that of the Whisper large-v3 encoder, and both models were quantized to 16-bit to ensure a fair comparison with the Qwen2.5-Omni audio encoder. As shown in Table~\ref{tab:performance}, DESAMO achieved the highest intent classification accuracy, outperforming both baselines while using the smallest total model size involved in inference. We observed that ASR errors in cascaded systems often led to intent misclassification, as illustrated by the examples in Fig.~\ref{fig:result}. In contrast, the Audio LLM in DESAMO correctly handled these cases by avoiding reliance on intermediate transcripts. Since the Fluent Speech Commands dataset is relatively clean in terms of pronunciation and recording conditions, we expect this performance gap to widen further in more realistic and noisy environments.

As a side note, we found that the system also handled non-English voice commands without any modification of the prompt, aligning with the cross-lingual generalization patterns observed in prior work~\cite{wang2023blsp}.

\section{Conclusion and Future Work}

We present DESAMO, an embedded smart home system that uses Audio LLMs to support elderly users through natural voice-based control and passive monitoring of critical ambient events, offering robust interaction entirely on edge hardware without compromising privacy. Beyond elderly care, this approach opens up possibilities for broader applications of edge-based Audio LLMs in contexts where privacy or network access are constrained. For future work, we plan to extend the system to multimodal settings by utilizing visual input for richer context understanding and to optimize inference latency --- currently around 5.3 seconds in headless mode --- for more responsive on-device interaction.

%%
%% The acknowledgments section is defined using the "acks" environment
%% (and NOT an unnumbered section). This ensures the proper
%% identification of the section in the article metadata, and the
%% consistent spelling of the heading.
\begin{acks}
This research was supported by National IT Industry Promotion Agency (NIPA) grant funded by the Korea government (MSIT) (Project Name: Development and demonstration of a fully autonomous speed sprayer (SS) based on on-device AI, Project Number: PJT-25-033312) and Culture, Sports and Tourism R\&D Program through the Korea Creative Content Agency grant funded by the Ministry of Culture, Sports and Tourism in 2023 (Project Name: Development of personalized exhibition viewing concierge service technology for the visually impaired, Project Number: RS-2023-00303777, Contribution Rate: 50\%).

\end{acks}

%%
%% The next two lines define the bibliography style to be used, and
%% the bibliography file.
\bibliographystyle{ACM-Reference-Format}
\bibliography{reference}

%%% -*-BibTeX-*-
%%% Do NOT edit. File created by BibTeX with style
%%% ACM-Reference-Format-Journals [18-Jan-2012].

\begin{thebibliography}{19}

%%% ====================================================================
%%% NOTE TO THE USER: you can override these defaults by providing
%%% customized versions of any of these macros before the \bibliography
%%% command.  Each of them MUST provide its own final punctuation,
%%% except for \shownote{} and \showURL{}.  The latter two
%%% do not use final punctuation, in order to avoid confusing it with
%%% the Web address.
%%%
%%% To suppress output of a particular field, define its macro to expand
%%% to an empty string, or better, \unskip, like this:
%%%
%%% \newcommand{\showURL}[1]{\unskip}   % LaTeX syntax
%%%
%%% \def \showURL #1{\unskip}           % plain TeX syntax
%%%
%%% ====================================================================

\ifx \showCODEN    \undefined \def \showCODEN     #1{\unskip}     \fi
\ifx \showISBNx    \undefined \def \showISBNx     #1{\unskip}     \fi
\ifx \showISBNxiii \undefined \def \showISBNxiii  #1{\unskip}     \fi
\ifx \showISSN     \undefined \def \showISSN      #1{\unskip}     \fi
\ifx \showLCCN     \undefined \def \showLCCN      #1{\unskip}     \fi
\ifx \shownote     \undefined \def \shownote      #1{#1}          \fi
\ifx \showarticletitle \undefined \def \showarticletitle #1{#1}   \fi
\ifx \showURL      \undefined \def \showURL       {\relax}        \fi
% The following commands are used for tagged output and should be
% invisible to TeX
\providecommand\bibfield[2]{#2}
\providecommand\bibinfo[2]{#2}
\providecommand\natexlab[1]{#1}
\providecommand\showeprint[2][]{arXiv:#2}

\bibitem[Chen et~al\mbox{.}(2024a)]%
        {chen2024octopus}
\bibfield{author}{\bibinfo{person}{Wei Chen}, \bibinfo{person}{Zhiyuan Li}, {and} \bibinfo{person}{Mingyuan Ma}.} \bibinfo{year}{2024}\natexlab{a}.
\newblock \showarticletitle{Octopus: On-device language model for function calling of software APIs}.
\newblock \bibinfo{journal}{\emph{arXiv e-prints}} (\bibinfo{year}{2024}), \bibinfo{pages}{arXiv--2404}.
\newblock


\bibitem[Chen et~al\mbox{.}(2024b)]%
        {chen2024voicebench}
\bibfield{author}{\bibinfo{person}{Yiming Chen}, \bibinfo{person}{Xianghu Yue}, \bibinfo{person}{Chen Zhang}, \bibinfo{person}{Xiaoxue Gao}, \bibinfo{person}{Robby~T Tan}, {and} \bibinfo{person}{Haizhou Li}.} \bibinfo{year}{2024}\natexlab{b}.
\newblock \showarticletitle{Voicebench: Benchmarking llm-based voice assistants}.
\newblock \bibinfo{journal}{\emph{arXiv preprint arXiv:2410.17196}} (\bibinfo{year}{2024}).
\newblock


\bibitem[Chu et~al\mbox{.}(2023)]%
        {chu2023qwen}
\bibfield{author}{\bibinfo{person}{Yunfei Chu}, \bibinfo{person}{Jin Xu}, \bibinfo{person}{Xiaohuan Zhou}, \bibinfo{person}{Qian Yang}, \bibinfo{person}{Shiliang Zhang}, \bibinfo{person}{Zhijie Yan}, \bibinfo{person}{Chang Zhou}, {and} \bibinfo{person}{Jingren Zhou}.} \bibinfo{year}{2023}\natexlab{}.
\newblock \showarticletitle{Qwen-audio: Advancing universal audio understanding via unified large-scale audio-language models}.
\newblock \bibinfo{journal}{\emph{arXiv preprint arXiv:2311.07919}} (\bibinfo{year}{2023}).
\newblock


\bibitem[Dawood et~al\mbox{.}(2024)]%
        {dawood2024management}
\bibfield{author}{\bibinfo{person}{Rusul~F. Dawood}, \bibinfo{person}{Mahmood~B. Mahmood}, {and} \bibinfo{person}{Rahma~S. Alsawaf}.} \bibinfo{year}{2024}\natexlab{}.
\newblock \showarticletitle{Management of Smart Home Using the Internet of Things: A Review}.
\newblock \bibinfo{journal}{\emph{Scientific Research Journal of Engineering and Computer Sciences}} \bibinfo{volume}{4}, \bibinfo{number}{1} (\bibinfo{date}{January} \bibinfo{year}{2024}), \bibinfo{pages}{1--8}.
\newblock


\bibitem[Everson et~al\mbox{.}(2024)]%
        {everson2024towards}
\bibfield{author}{\bibinfo{person}{Kevin Everson}, \bibinfo{person}{Yile Gu}, \bibinfo{person}{Huck Yang}, \bibinfo{person}{Prashanth~Gurunath Shivakumar}, \bibinfo{person}{Guan-Ting Lin}, \bibinfo{person}{Jari Kolehmainen}, \bibinfo{person}{Ivan Bulyko}, \bibinfo{person}{Ankur Gandhe}, \bibinfo{person}{Shalini Ghosh}, \bibinfo{person}{Wael Hamza}, {et~al\mbox{.}}} \bibinfo{year}{2024}\natexlab{}.
\newblock \showarticletitle{Towards ASR robust spoken language understanding through in-context learning with word confusion networks}. In \bibinfo{booktitle}{\emph{ICASSP 2024-2024 IEEE International Conference on Acoustics, Speech and Signal Processing (ICASSP)}}. IEEE, \bibinfo{pages}{12856--12860}.
\newblock


\bibitem[Fathullah et~al\mbox{.}(2024)]%
        {fathullah2024audiochatllama}
\bibfield{author}{\bibinfo{person}{Yassir Fathullah}, \bibinfo{person}{Chunyang Wu}, \bibinfo{person}{Egor Lakomkin}, \bibinfo{person}{Ke Li}, \bibinfo{person}{Junteng Jia}, \bibinfo{person}{Yuan Shangguan}, \bibinfo{person}{Jay Mahadeokar}, \bibinfo{person}{Ozlem Kalinli}, \bibinfo{person}{Christian Fuegen}, {and} \bibinfo{person}{Mike Seltzer}.} \bibinfo{year}{2024}\natexlab{}.
\newblock \showarticletitle{AudioChatLlama: Towards General-Purpose Speech Abilities for LLMs}. In \bibinfo{booktitle}{\emph{Proceedings of the 2024 Conference of the North American Chapter of the Association for Computational Linguistics: Human Language Technologies (Volume 1: Long Papers)}}. \bibinfo{pages}{5522--5532}.
\newblock


\bibitem[Ghosh et~al\mbox{.}(2024)]%
        {ghosh2024gama}
\bibfield{author}{\bibinfo{person}{Sreyan Ghosh}, \bibinfo{person}{Sonal Kumar}, \bibinfo{person}{Ashish Seth}, \bibinfo{person}{Chandra Kiran~Reddy Evuru}, \bibinfo{person}{Utkarsh Tyagi}, \bibinfo{person}{S Sakshi}, \bibinfo{person}{Oriol Nieto}, \bibinfo{person}{Ramani Duraiswami}, {and} \bibinfo{person}{Dinesh Manocha}.} \bibinfo{year}{2024}\natexlab{}.
\newblock \showarticletitle{GAMA: A Large Audio-Language Model with Advanced Audio Understanding and Complex Reasoning Abilities}. In \bibinfo{booktitle}{\emph{Proceedings of the 2024 Conference on Empirical Methods in Natural Language Processing}}. \bibinfo{pages}{6288--6313}.
\newblock


\bibitem[Guam{\'a}n et~al\mbox{.}(2018)]%
        {guaman2018device}
\bibfield{author}{\bibinfo{person}{Steven Guam{\'a}n}, \bibinfo{person}{Adri{\'a}n Calvopi{\~n}a}, \bibinfo{person}{Pamela Orta}, \bibinfo{person}{Freddy Tapia}, {and} \bibinfo{person}{Sang~Guun Yoo}.} \bibinfo{year}{2018}\natexlab{}.
\newblock \showarticletitle{Device control system for a smart home using voice commands: A practical case}. In \bibinfo{booktitle}{\emph{Proceedings of the 2018 10th International Conference on Information Management and Engineering}}. \bibinfo{pages}{86--89}.
\newblock


\bibitem[Lugosch et~al\mbox{.}(2019)]%
        {lugosch2019speech}
\bibfield{author}{\bibinfo{person}{Loren Lugosch}, \bibinfo{person}{Mirco Ravanelli}, \bibinfo{person}{Patrick Ignoto}, \bibinfo{person}{Vikrant~Singh Tomar}, {and} \bibinfo{person}{Yoshua Bengio}.} \bibinfo{year}{2019}\natexlab{}.
\newblock \showarticletitle{Speech Model Pre-Training for End-to-End Spoken Language Understanding}. In \bibinfo{booktitle}{\emph{Proc. Interspeech 2019}}. \bibinfo{pages}{814--818}.
\newblock


\bibitem[Mitra et~al\mbox{.}(2019)]%
        {mitra2019leveraging}
\bibfield{author}{\bibinfo{person}{Vikramjit Mitra}, \bibinfo{person}{Sue Booker}, \bibinfo{person}{Erik Marchi}, \bibinfo{person}{David~Scott Farrar}, \bibinfo{person}{Ute~Dorothea Peitz}, \bibinfo{person}{Bridget Cheng}, \bibinfo{person}{Ermine Teves}, \bibinfo{person}{Anuj Mehta}, {and} \bibinfo{person}{Devang Naik}.} \bibinfo{year}{2019}\natexlab{}.
\newblock \showarticletitle{Leveraging Acoustic Cues and Paralinguistic Embeddings to Detect Expression from Voice}. In \bibinfo{booktitle}{\emph{Proc. Interspeech 2019}}. \bibinfo{pages}{1651--1655}.
\newblock


\bibitem[Patil et~al\mbox{.}(2024)]%
        {patil2024gorilla}
\bibfield{author}{\bibinfo{person}{Shishir~G Patil}, \bibinfo{person}{Tianjun Zhang}, \bibinfo{person}{Xin Wang}, {and} \bibinfo{person}{Joseph~E Gonzalez}.} \bibinfo{year}{2024}\natexlab{}.
\newblock \showarticletitle{Gorilla: Large language model connected with massive apis}.
\newblock \bibinfo{journal}{\emph{Advances in Neural Information Processing Systems}}  \bibinfo{volume}{37} (\bibinfo{year}{2024}), \bibinfo{pages}{126544--126565}.
\newblock


\bibitem[Radford et~al\mbox{.}(2023)]%
        {radford2023robust}
\bibfield{author}{\bibinfo{person}{Alec Radford}, \bibinfo{person}{Jong~Wook Kim}, \bibinfo{person}{Tao Xu}, \bibinfo{person}{Greg Brockman}, \bibinfo{person}{Christine McLeavey}, {and} \bibinfo{person}{Ilya Sutskever}.} \bibinfo{year}{2023}\natexlab{}.
\newblock \showarticletitle{Robust speech recognition via large-scale weak supervision}. In \bibinfo{booktitle}{\emph{Proceedings of the 40th International Conference on Machine Learning}}. \bibinfo{pages}{28492--28518}.
\newblock


\bibitem[Schick et~al\mbox{.}(2023)]%
        {schick2023toolformer}
\bibfield{author}{\bibinfo{person}{Timo Schick}, \bibinfo{person}{Jane Dwivedi-Yu}, \bibinfo{person}{Roberto Dess{\`\i}}, \bibinfo{person}{Roberta Raileanu}, \bibinfo{person}{Maria Lomeli}, \bibinfo{person}{Eric Hambro}, \bibinfo{person}{Luke Zettlemoyer}, \bibinfo{person}{Nicola Cancedda}, {and} \bibinfo{person}{Thomas Scialom}.} \bibinfo{year}{2023}\natexlab{}.
\newblock \showarticletitle{Toolformer: Language models can teach themselves to use tools}.
\newblock \bibinfo{journal}{\emph{Advances in Neural Information Processing Systems}}  \bibinfo{volume}{36} (\bibinfo{year}{2023}), \bibinfo{pages}{68539--68551}.
\newblock


\bibitem[Schomakers et~al\mbox{.}(2020)]%
        {schomakers2020privacy}
\bibfield{author}{\bibinfo{person}{Eva‑Maria Schomakers}, \bibinfo{person}{Hannah Biermann}, {and} \bibinfo{person}{Martina Ziefle}.} \bibinfo{year}{2020}\natexlab{}.
\newblock \showarticletitle{Understanding Privacy and Trust in Smart Home Environments}. In \bibinfo{booktitle}{\emph{HCII 2020 (LNCS)}}. \bibinfo{pages}{513--532}.
\newblock


\bibitem[Tang et~al\mbox{.}(2024)]%
        {tang2024extending}
\bibfield{author}{\bibinfo{person}{Changli Tang}, \bibinfo{person}{Wenyi Yu}, \bibinfo{person}{Guangzhi Sun}, \bibinfo{person}{Xianzhao Chen}, \bibinfo{person}{Tian Tan}, \bibinfo{person}{Wei Li}, \bibinfo{person}{Lu Lu}, \bibinfo{person}{Zejun Ma}, {and} \bibinfo{person}{Chao Zhang}.} \bibinfo{year}{2024}\natexlab{}.
\newblock \showarticletitle{Extending large language models for speech and audio captioning}. In \bibinfo{booktitle}{\emph{ICASSP 2024-2024 IEEE International Conference on Acoustics, Speech and Signal Processing (ICASSP)}}. IEEE, \bibinfo{pages}{11236--11240}.
\newblock


\bibitem[Vipperla et~al\mbox{.}(2008)]%
        {vipperla2008longitudinal}
\bibfield{author}{\bibinfo{person}{Rama Vipperla}, \bibinfo{person}{Steve Renals}, {and} \bibinfo{person}{Joe Frankel}.} \bibinfo{year}{2008}\natexlab{}.
\newblock \showarticletitle{Longitudinal Study of ASR Performance on Ageing Voices}. In \bibinfo{booktitle}{\emph{Proceedings of Interspeech 2008}}. \bibinfo{pages}{2550--2553}.
\newblock


\bibitem[Wang et~al\mbox{.}(2023)]%
        {wang2023blsp}
\bibfield{author}{\bibinfo{person}{Chen Wang}, \bibinfo{person}{Minpeng Liao}, \bibinfo{person}{Zhongqiang Huang}, \bibinfo{person}{Jinliang Lu}, \bibinfo{person}{Junhong Wu}, \bibinfo{person}{Yuchen Liu}, \bibinfo{person}{Chengqing Zong}, {and} \bibinfo{person}{Jiajun Zhang}.} \bibinfo{year}{2023}\natexlab{}.
\newblock \showarticletitle{Blsp: Bootstrapping language-speech pre-training via behavior alignment of continuation writing}.
\newblock \bibinfo{journal}{\emph{arXiv preprint arXiv:2309.00916}} (\bibinfo{year}{2023}).
\newblock


\bibitem[Xu et~al\mbox{.}(2025)]%
        {xu2025qwen2}
\bibfield{author}{\bibinfo{person}{Jin Xu}, \bibinfo{person}{Zhifang Guo}, \bibinfo{person}{Jinzheng He}, \bibinfo{person}{Hangrui Hu}, \bibinfo{person}{Ting He}, \bibinfo{person}{Shuai Bai}, \bibinfo{person}{Keqin Chen}, \bibinfo{person}{Jialin Wang}, \bibinfo{person}{Yang Fan}, \bibinfo{person}{Kai Dang}, {et~al\mbox{.}}} \bibinfo{year}{2025}\natexlab{}.
\newblock \showarticletitle{Qwen2. 5-omni technical report}.
\newblock \bibinfo{journal}{\emph{arXiv preprint arXiv:2503.20215}} (\bibinfo{year}{2025}).
\newblock


\bibitem[Yue and Ping(2017)]%
        {yue2017voice}
\bibfield{author}{\bibinfo{person}{Chan~Zhen Yue} {and} \bibinfo{person}{Shum Ping}.} \bibinfo{year}{2017}\natexlab{}.
\newblock \showarticletitle{Voice activated smart home design and implementation}. In \bibinfo{booktitle}{\emph{2017 2nd International Conference on Frontiers of Sensors Technologies (ICFST)}}. IEEE, \bibinfo{pages}{489--492}.
\newblock


\end{thebibliography}

%%
%% If your work has an appendix, this is the place to put it.

\end{document}